\documentclass[conference]{IEEEtran}
\IEEEoverridecommandlockouts
\usepackage{cite}
\usepackage{amsmath,amssymb,amsfonts}
\usepackage{algorithmic}
\usepackage{textcomp}

\def\BibTeX{{\rm B\kern-.05em{\sc i\kern-.025em b}\kern-.08em
    T\kern-.1667em\lower.7ex\hbox{E}\kern-.125emX}}
\usepackage{ifpdf}
\usepackage{cite}

\ifCLASSINFOpdf
\usepackage[pdftex]{graphicx}
\graphicspath{{../pdf/}{../jpeg/}{./pic/}}
\DeclareGraphicsExtensions{.pdf,.jpeg,.png}
\else
\usepackage[dvips]{graphicx}
\graphicspath{{../eps/}}
\DeclareGraphicsExtensions{.eps}
\fi

\usepackage{amsmath}
\usepackage{algorithmic}
\usepackage{array}
\usepackage{pgfplots} 
\usepackage[para]{footmisc}

\usepackage[font=normalsize,labelfont=sf,textfont=sf]{caption}
\usepackage[font=normalsize,labelfont=sf,textfont=sf]{subcaption}

\usepackage{lmodern}  
\usepackage{mathrsfs} 

\usepackage{amssymb}  
\usepackage{csquotes}

\usepackage[ruled,vlined]{algorithm2e}
\usepackage{mathtools} 
\usepackage{cases} 
\usepackage{empheq}
\usepackage[capitalize]{cleveref} 
\usepackage{cuted} 
\usepackage{bm}
\usepackage{amsthm}

\usepackage{tabulary,booktabs} 
\usepackage[flushleft]{threeparttable}
\usepackage{booktabs,caption}

\usepackage{tensor}
\usepackage{multirow}

\usepackage{tabularx}
\usepackage{booktabs}
\usepackage[]{xcolor}
\usepackage[labelfont=bf,format=plain,justification=raggedright,singlelinecheck=false]{caption}
\usepackage[hyphens]{url}
\expandafter\def\expandafter\UrlBreaks\expandafter{\UrlBreaks
  \do\a\do\b\do\c\do\d\do\e\do\f\do\g\do\h\do\i\do\j%
  \do\k\do\l\do\m\do\n\do\o\do\p\do\q\do\r\do\s\do\t%
  \do\u\do\v\do\w\do\x\do\y\do\z\do\A\do\B\do\C\do\D%
  \do\E\do\F\do\G\do\H\do\I\do\J\do\K\do\L\do\M\do\N%
  \do\O\do\P\do\Q\do\R\do\S\do\T\do\U\do\V\do\W\do\X%
  \do\Y\do\Z}

\theoremstyle{definition}
\newtheorem{definition}{Definition}

\newcommand{\E}{\mathrm{E}}

\newcommand{\bb}[1]{\mathbb{#1}}

\newcommand{\mcal}[1]{\mathcal{#1}}

\newcommand{\C}{C(J,\bb{R}^L)}

\usepackage{listings}
\usepackage{romannum}

\begin{document}

\title{
	A Dynamic Resource Allocation Framework for Synchronizing Metaverse with IoT Service and Data \\
}

\author
{\IEEEauthorblockN{
Yue~Han\IEEEauthorrefmark{1}\IEEEauthorrefmark{2}\IEEEauthorrefmark{3}, 
Dusit~Niyato\IEEEauthorrefmark{1}
Cyril~Leung\IEEEauthorrefmark{4}\IEEEauthorrefmark{5},
Chunyan~Miao\IEEEauthorrefmark{1}\IEEEauthorrefmark{5},
Dong~In~Kim\IEEEauthorrefmark{6}
}

\IEEEauthorblockA{
\IEEEauthorrefmark{2} Alibaba-NTU JRI, Interdisciplinary
Graduate School, Nanyang Technological University (NTU)
\IEEEauthorrefmark{1} SCSE, NTU \\
\IEEEauthorrefmark{3} Alibaba Group
\IEEEauthorrefmark{5} LILY, NTU
\IEEEauthorrefmark{4} ECE, The University of British Columbia
\IEEEauthorrefmark{6} ECE, Sungkyunkwan University\\
}
}

\maketitle
\begin{abstract}
Spurred by the severe restrictions on mobility due to the COVID-19 pandemic, there is currently intense interest in developing the Metaverse, to offer virtual services/business online. A key enabler of such virtual service is the digital twin, i.e., a digital replication of real-world entities in the Metaverse, e.g., city twin, avatars, etc. The real-world data collected by IoT devices and sensors are key for synchronizing the two worlds. In this paper, we consider the scenario in which a group of IoT devices are employed by the Metaverse platform to collect such data on behalf of virtual service providers (VSPs). Device owners, who are self-interested, dynamically select a VSP to maximize rewards. We adopt hybrid evolutionary dynamics, in which heterogeneous device owner populations can employ different revision protocols to update their strategies. Extensive simulations demonstrate that a hybrid protocol can lead to evolutionary stable states. 
\end{abstract}


\section{Introduction}
{COVID-19} has dramatically changed work and life styles, from physical to online/virtual experience, e.g., attending virtual job fair, concert, meeting, and graduation ceremony. Generally, those virtual spaces are referred to as 
`Metaverse', where `meta' means virtual and transcendence, and `verse' represents universe. Formally, Metaverse describes a  `{computer-generated, multi-user, three-dimensional interfaces in which users experience other participants as being present in the environment}' \cite{schroederSocialInteractionVirtual2002}.

To realize the Metaverse, digitization of the real world is needed. 
In the Metaverse, {information} measured in bits are the main existence. 
Living and nonliving entities in the physical world, e.g., drivers, roads, can be \textit{scanned} by smart devices, e.g., IoT devices and sensors, and be digitally replicated in the Metaverse. This digital replication idea has been recently studied by many researchers, around a concept called \textit{digital twin} \cite{petrova-antonovaDigitalTwinModeling2020,fullerDigitalTwinEnabling2020,elsaddikDigitalTwinsConvergence2018}. An actual application is the smart city twin \cite{petrova-antonovaDigitalTwinModeling2020}, in which a virtual city could be operated in the Metaverse, so that new polices, e.g., traffic intervention, could be $A/B$ tested in the Metaverse prior to implementation in the physical world. The benefits of having digital twins in the Metaverse are obvious, as the Metaverse can facilitate the monitoring, understanding, and optimization of the real-world business and its functions with infinite, cost-efficient experiments, collecting continuous feedback without affecting the business in the physical world. The benefits are particularly important when the test of the intervention is expensive and irreversible in the real world, e.g., potential environment destruction.

\begin{figure*}[]
\begin{minipage}[b]{0.64\linewidth}
\centering
\includegraphics[width=\linewidth]{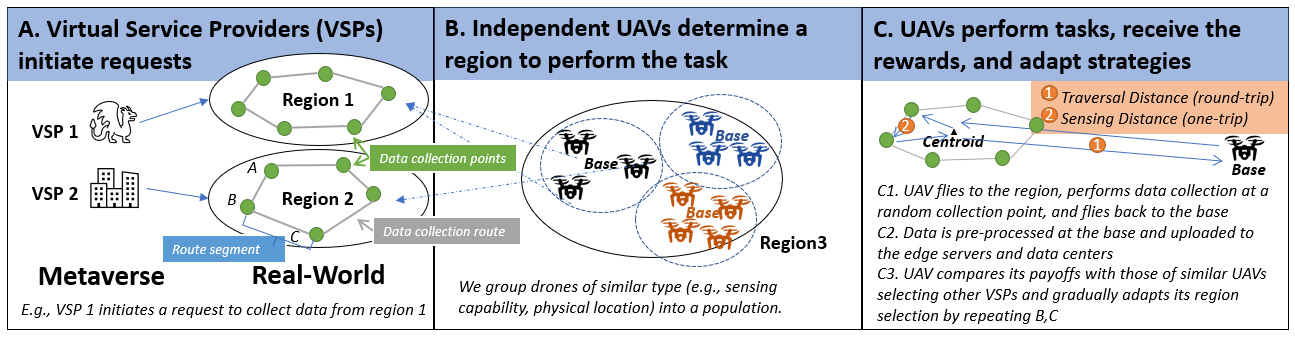}
\caption{System Model: IoT-assisted data collection to enable sync between the Metaverse and the physical world. Here, IoT is exemplified by UAV.}
\label{fig:system4}
\end{minipage} \quad
\begin{minipage}[b]{0.34\linewidth}
\centering
\includegraphics[width=\linewidth]{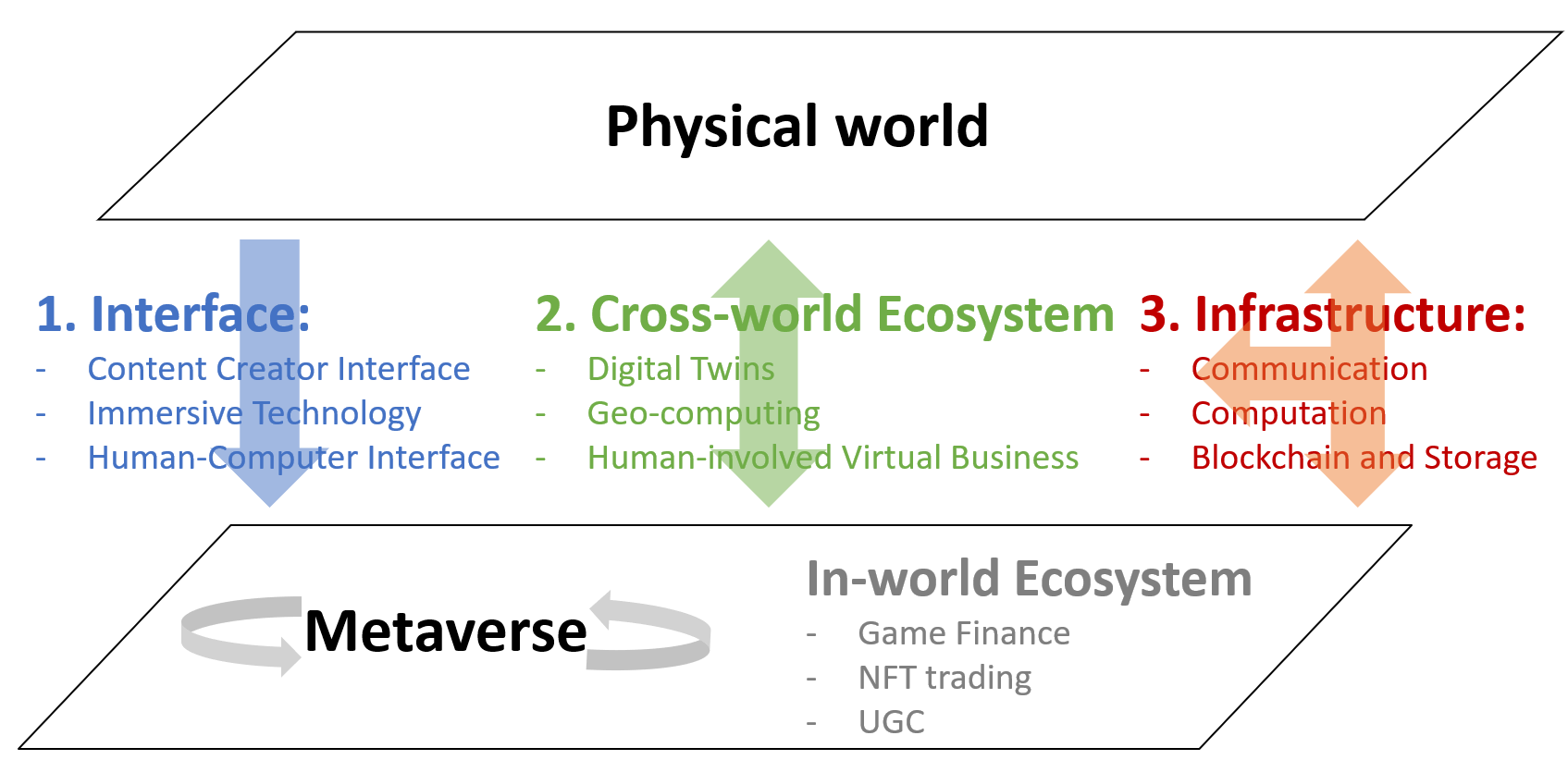}
\caption{Four Components for the Metaverse}
\label{fig:arc}
\end{minipage} 
\end{figure*}

A key aspect in implementing a better digital twin is the continuous data synchronization(sync) in the Metaverse, i.e., acquiring fresh data from the physical world to keep the digital twin updated in real time. For example, a virtual driver training service provider \cite{taheriVirtualRealityDriving2017}, which simulates the virtual road using real data (such as road, drivers, traffic, and weather), needs to have continuous data updates to ensure that its service (e.g., driving skill test) is realistic, and the trainee's real driving skill can be assessed. 
To enable such frequent sync across two worlds, we consider to employ IoT devices, such as autonomous driving cars, unmanned aerial vehicle (UAV), and smart phones, to collect the data around a region that a virtual service provider (VSP) is of interest.

In general, we consider a resource allocation problem to support the sync between the Metaverse and the real world. Our system model is shown in \cref{fig:system4}. Although we use UAVs in this illustrative example, the model is general and can be extended to different types of IoT devices. Suppose a VSP is interested in collecting data from a region in the real world (e.g., Region $1$ in \cref{fig:system4}) for a service offering, e.g., simulating a road. The VSP first announces a data collection task (step ${A}$) over a particular region for a particular period, e.g., a week. Then, the Metaverse platform, which hosts multiple VSPs, supports the VSP in its synchronizing task by engaging the services of IoT devices in the real world 
(e.g., services of the UAVs located in a nearby Region $3$ in \cref{fig:system4}).

Each UAV is owned by an independent entity that can decide which task to work on (step $B$). After a UAV collects the sensing data, data pre-processing takes place at each UAV's base (e.g., home), following which the processed data are transmitted to 1) the local Metaverse platform cached at the edge clouds to support real-time and interactive services, and 2) the core of the Metaverse platform located at data centers with large capacity of computation, communication, and storage, supporting the Metaverse's core functionality, e.g. intensive simulations of the virtual worlds, continuous user identity. 

As the duration of a task can last relative long, a UAV owner can gradually improve its decision making by observing the reward obtained by similar UAVs selecting other sync tasks and learning from them (step $C$). Finally, an equilibrium state at which no UAV owner would unilaterally change its decision can be reached. 

The main contributions of this paper are as follows:
\begin{itemize}
\item We identify a resource allocation problem for virtual service sync selection in which the IoT devices in a particular region in the real world are hired by the Metaverse platform to support its hosted VSPs towards a more efficient and collaborative environment of virtual content creation.  
\item In consideration of a large number of independent IoT device owners at the edge and their self-interested nature, the equilibrium knowledge (full rationale) is hardly-achievable. We propose a dynamic approach founded on evolutionary game theory, in which the \textit{bounded-rational} device owners can gradually adapt their strategies towards the equilibrium state. 
\item Our evolutionary game framework considers a general sensing model and reward allocation scheme and thus can be extended to specific tasks that support the sync between the Metaverse and the real-world.  
\end{itemize}

\section{Metaverse Preliminaries and Related Work}
In this section, we first introduce the architecture that composes a Metaverse, followed by related works and the research gap. 

\subsection{Metaverse General Architecture}\label{subsection:metaverse-general-architecture}
Architectures such as \cite{radoffMetaverseValueChain2021} and \cite{duanMetaverseSocialGood2021} have been proposed for the Metaverse. Based on these, we propose a four-component architecture as shown in \cref{fig:arc}. It includes infrastructure, interface, cross-world ecosystem, and in-world ecosystem.

\textit{\textbf{Infrastructure}} (including communication, computation, blockchain, and other decentralization techniques) is to support all operations in the {physical world}, the {Metaverse}, and the connection between the two worlds. 
The services that inter-connect the two worlds are further grouped by (\romannum{1}) \textit{\textbf{Interface}}, which is related to immersive technologies to enrich human's subjective sense, and (\romannum{2}) \textit{\textbf{Cross-worlds Ecosystem}}, which consists of a variety of services to achieve the \textit{convergence} between the two worlds, e.g. digital city twin simulation with real-time IoT data \cite{petrova-antonovaDigitalTwinModeling2020} and virtual concert with real-time human performers\cite{kusumaChildrenVirtualConcert2020}. In particular, the resource allocation problem in the cross-world ecosystem, i.e., virtual service sync selection by IoT device owners, is the main focus of this paper. Finally, the Metaverse-enabled business and economics, e.g., the non-fungible token (NFT) trading, are categorized as \textit{\textbf{In-world Ecosystem}}, as it does not requires frequent data traversal across the two worlds. 

\subsection{Related Works}
Recently, due to the COVID-19 pandemic mobility restriction and the marketing by big tech companies such as Facebook and Microsoft \cite{brownBigTechWants2021}, the necessity to function in the virtual world has been demonstrated in various aspects of life. As such, 
the topic of Metaverse has attracted a lot of research attention, for its potential in service offerings in retailing \cite{gadallaMetaverseretailServiceQuality2013}, gaming \cite{volkCocreativeGameDevelopment2008}, education \cite{diazVirtualWorldResource2020},  and social-networking \cite{schroederSocialInteractionVirtual2002}. Some researchers have proposed to build prototypes to better understand the Metaverse and its economics, e.g., a university campus prototype \cite{duanMetaverseSocialGood2021} to study social good in the Metaverse. These examples illustrate the importance of convergence between the Metaverse and the physical world. 

\subsection{Research Gap: Resource Allocation in Converging the Metaverse and the Physical Worlds}
As the study of the Metaverse is still in its nascent stage,  
the resource allocation problem for syncing Metaverse with the assistance of IoT sensing data has not received much attention. Previous works, such as \cite{monetaArchitectureHeritageMetaverse2020,dionisio3DVirtualWorlds2013c,freySolipsisDecentralizedArchitecture2008}, mostly present a general picture of the field, summarizing technologies, architectures, and challenges involved in developing the Metaverse. Other works, such as \cite{leeSelfconfigurableLargescaleVirtual2011, tamaiConstructingSituatedLearning2011}, primarily focus on human-machine interaction aspects of the virtual learning experience (i.e., the interface component in our architecture). In contrast, our work is mainly focused on the resource allocation challenge to achieve convergence between the Metaverse and the physical world (i.e., the cross-world component in our architecture). 

\section{System Model and Problem Formulation}
\begin{table}[] 
\begin{threeparttable}
\caption{Notation Used in the System Model}
\label{tab:my-table}
\begin{tabular}{|p{1.3cm}|p{6.5cm}|}
\hline
\textbf{\textit{Notation}} & \textbf{\textit{Description}} \\ \hline
\specialrule{1pt}{0pt}{0pt}
$m,i,j$               & index of virtual service provider (VSP) and its target region                  \\ \hline
$p,q$               & index of IoT device population                     \\ \hline
$R_m$           &  reward                 \\ \hline
$D_m$           & total Euclidean distance of the sensing route                 \\ \hline
$d_m$         & length of the segmented sensing distance   \\ \hline
$N^p$                &  size of UAV population  $p$                  \\ \hline
$\mcal{S}^p$ & set of $S^p$ (pure) strategies \\ \hline
$\zeta^p$           & unit energy cost                     \\ \hline
$\eta^p_1,\eta^p_2$& power parameters\\ \hline
 $v^p,u^p$ &average velocity during traversal and sensing stage\\ \hline

$l^p_m$         & traversal distance from the base in population $p$ to the data collection point in region $m$  \\ \hline             
$x^p_m$         & strategy distribution, population state                    \\ \hline
$b^p_m$           & sensing data quality                     \\ \hline
$E_{s}^{m,p}$                & total energy consumption for traversal and sensing                    \\ \hline
$R_m^p$                &  received  reward              \\ \hline
$F^p_m(\bm{x}),\pi^p_m$                &  net payoff           \\ \hline 
\end{tabular}
\begin{tablenotes}
      \small
      \item Note that superscript and subscript are to denote the index of population and strategy respectively. To avoid confusion, parenthesis is applied when involving exponents, e.g., $(v^p)^3$.
    \end{tablenotes}
\end{threeparttable}
\end{table}

We consider a network consisting of a set $\mcal{M}=\{1,\ldots,m, \ldots, M\}$ of $M$ VSPs, each of which aims to collect data (e.g., geo-spatial data) from a target region in the physical world in order to sync its virtual services (e.g., digital twin of a road for virtual driver training) in the Metaverse. IoT devices, such as UAVs owned by individuals in a nearby region, are employed by the Metaverse platform to collect data for VSPs to use.
For a target region $m$, the VSPs or the Metaverse platform designs a data collection route covering the region (represented by the the grey line in \cref{fig:system4}), which can be further segmented based on the number of IoT devices that sense the region. 

For simplicity, we assume that the allocation of data collection points in region $m$ to the set of UAVs which selecting this region is uniform. Along a segmented sensing route (e.g., line segment $BC$ in \cref{fig:system4}A), a UAV senses the nearby environment and collects the data. Assume that the data collection tasks last relatively long, e.g., one week, during which time UAVs collect data multiple times.

Let $\mcal{N}=\{1,\ldots,n,\ldots N\}$ denote a set of $N$ IoT devices, exemplified by UAVs, owned by $N$ independent individuals located in a nearby region (e.g., Region $3$ in \cref{fig:system4}B). UAV devices located at in their bases, such as owners' homes or community centers, need to fly to a region's, e.g., region $m$, data collection point to perform the task and return back for charging. We refer this two-way travel distance as the \textit{traversal distance}. As the traversal distance depends on which data collection point is assigned, we will approximate its value by the distance between the base and the center of region $m$.

Presented with $M$ VSPs' sync task requests, an UAV device owner needs to make a decision on which VSP to work for. Given that there can be a large number of UAV device owners and the full rational (equilibrium knowledge) among them is hard to achieve, a device owner can make irrational one-off decision. Since the sync task lasts several rounds, an owner can gradually improve his decisions towards equilibrium ones by observing the payoffs received by nearby UAVs with the same type of tasks.

In summary, there are two stages in our system model:
\begin{enumerate}
\item \textbf{Population Game Formulation}: Nearby UAVs with similar types (e.g., sensing capability, unit energy cost) are regarded as anonymously same and communicable and are grouped into the same population. 
Intra-population communication can be achieved in a \textit{pairwise} way (e.g., device-to-device), a \textit{centralized} approach (e.g., requesting the payoff information from the BS or checking the bulletin board, a place to publish virtual task \cite{duanMetaverseSocialGood2021}, in the Metaverse platform), or a combination.
Information checking via a centralized approach is costly and prone to delay but yields a complete picture of the payoffs. In contrast, the pairwise approach is cheap and fast but provides only limited payoff information among the interacting UAVs.

\item \textbf{Hybrid Evolutionary Dynamics}:
UAV populations may have different frequency to switch to a more costly centralized approach. Thus UAVs populations may have heterogeneous knowledge of the payoff information, leading to different strategy adjustments per round. To address this, we formulate the problem as a \textit{heterogeneous multi-population game} and adopt \textit{hybrid evolutionary dynamics} \cite{sandholmPopulationGamesEvolutionary2011} (a generalization of pure replicator dynamics)  to solve for the evolutionary stable strategy (ESS). 

\end{enumerate}
\subsection{Population Formulation}
We group neighbor UAVs (e.g., in the same community) of similar type (characterized by e.g., sensing capacity, traversal costs, communication frequency with the BS) 
into a population, and let $\mcal{P}=\{1,\ldots,p,\ldots, P\}$ denote the set of populations. For any population $p$, its UAVs are regarded as anonymously same, e.g., with similar traversal distance $l^p_m$ to region $m$, unit energy cost $\zeta^p$, and sensing data quality $b^p_m$. Let $N^p$ denote the size of population $p$ so that $\sum_{p\in\mcal{P}}N^p=N$. We refer $\mcal{N}$, the set of $N$ UAV owners, to as the \textit{society}. 

Let $\mcal{S}^p=\{1,\ldots, s^p,\ldots, S^p\}$ denote the set of (pure) \textit{strategies}, i.e., there are $S^p$ VSP sync tasks, which can be selected by an owner in population $p$. A typical element in $\mcal{S}^p$ is denoted by $m,i,j$. 
The total number of pure strategies in all populations is denoted by $S=\sum_{p\in\mcal{P}} S^p$. 

The set of \textit{population states} (or \textit{strategy distribution}) for population $p$ is denoted by the simplex in $\mathbf{R}^{S^p}_+$, $\Delta^p=\{\bm{x}^p\in\mathbf{R}^{S^p}_+: \sum_{m\in \mcal{S}^p}x^p_m=1\}$, where $\mathbf{R}^{S^p}_+$ is the space of non-negative real-numbers of dimension $S^p$ and $x^p_m\in\mathbf{R}_+$ is the percentage of population $p$ selecting the region $m$. 

Define $\Theta=\prod_{p\in\mcal{P}}\Delta^p=\{\bm{x}=[\bm{x}^1,\ldots,\bm{x}^{P}]\in\mathbf{R}^S_+: \bm{x}^p\in \Delta^p\}$. The element of $\Theta$ describes the \textit{social states}, namely the joint behaviors in all $P$ populations at once. 

\subsection{UAV Sensing Model}
Let $D_m$ denote the total Euclidean distance of the sensing route over the region $m$, which is to be shared among the number of $\sum_{p\in\mcal{P}} x^p_m N^p$ UAVs selecting the region. We use the map $g^p_m: \mathbf{R}_+\rightarrow \mathbf{R}_{++}$ to refer to the route sharing policy, yielding the length of a segmented sensing route in region $m$ for a UAV in population $p$. For simplicity, we consider that the route is evenly divided among UAVs selecting the region. Therefore, the \textit{sensing distance} for a $p$-population UAV selecting region $m$ is defined by 
$d_m = g^p_m(D_m) = {D_m}/( {\sum_{q\in\mcal{P}} x^q_m N^q})$.
 
Following  \cite{zhangPredictiveDeploymentUAV2021,limFederatedLearningUAVEnabled2021}, each UAV's energy cost consists of three components: the propulsion power $\eta^p_1$, hover power $\eta^p_2$ during service stage, and the transmission power. As we consider that the UAV transmits data after flying back to its base, the costs of propulsion and hover are of most interest. Following \cite{zengEnergyMinimizationWireless2019}, the energy costs during a UAV's acceleration and deceleration stages are ignored, and the propulsion power are considered to be constant for a fixed flying speed. Thus, for a $p$-population UAV selecting region $m$, the total energy consumption for traversal and sensing distance is defined by:
\begin{equation*}\label{key}
E_{s}^{n,p}= \eta^p_1\frac{l^p_n}{v^p}+\eta^p_2\frac{d_m}{u^p} = 
 \eta^p_1\frac{l^p_n}{v^p}+\eta^p_2{\frac{D_m}{{u^p}\sum_{q\in\mcal{P}} x^q_m N^q}} ,
\end{equation*}
where $v^p$ and $u^p$ are the average flying speeds during the traversal stage and the sensing/hovering stage.

\subsection{UAV Utility Model}
Without loss of generality, we consider the reward given by VSP $m$ is evenly allocated to the UAVs working in the region, except for a weighting factor $b^p_m$ which represents the the quality of the sensor data, e.g., measured by the sampling rate \cite{tongDeepReinforcementLearning2020}. A higher (intensive) sampling rate leads to the sensor data of higher quality, capturing more objects per frame. Therefore, the reward received by a $p$-population UAV selecting region $m$ is defined by 
\begin{equation}\label{key}
R^p_m =  \frac{ b^p_m }{\sum_{q\in\mcal{P}} x^q_m N^q b^q_m }R_m.
\end{equation}

Let $\pi^p=[\pi^p_m]_{m\in \mcal{S}^p}$ denote the payoff vectors for population $p$. For any UAV in population $p$ selecting VSP $m$, its received utility $\pi^p_m$ is defined as follows:
\begin{equation} \label{eq:utility}
\pi^p_m= F^p_m(\bm{x})= R^p_m - \zeta^p E_{s}^{n,p} ,
\end{equation}
where $\zeta^p$ is the unit cost per Joule of energy.

\section{Heterogeneous Evolutionary Game}
A population game can be formulated as a set of players $\mcal{N}$, a set of populations $\mcal{P}$, a set of strategies $\mcal{S}^p\, (p\in\mcal{P})$, and a payoff function $F$. 
The definition and requirement of $F$ is as follows: $F:\Theta\rightarrow \mathbf{R}^S$ is a continuous map\footnote{Besides the continuous requirement, $F$ are imposed by stronger requirement of being Lipschitz continuous or continuously differentiable $(C^1)$.} assigning each social state $\bm{x}$ a vector of $S$ payoffs, one for each strategy in each population. Let $F^p_m:\Theta \rightarrow \mathbf{R}$ denote the payoff function for strategy $i\in S^p$, defined by \eqref{eq:utility}. Then, $F=[F^p]_{p\in\mcal{P}}$, where $F^p=[F^p_m]_{m\in \mcal{S}^p}$. 

The UAV owners' behaviors adaptation can be modeled by \textit{revision protocols}, which are defined as follows:
\begin{definition}A revision protocol in population $p$ is a map $\rho^p:\mathbf{R}^{S^p}\times \Delta^p \rightarrow \mathbf{R}^{S^p\times  \mcal{S}^p}_+$. The scalar $\rho^p_{m,i}(\pi^p,\bm{x}^p)$ defines the \textit{conditional switch rate} from strategy $m\in S^p$ to strategy $i\in \mcal{S}^p$ given payoff vector $\pi^p$ and population state $\bm{x}^p$. For convenience, we refer to the collection $\rho=[\rho^1,\ldots,\rho^P]$ as a revision protocol.
\end{definition}

Generally, a revision protocol $\rho$ defines a continuous-time evolutionary process over the populations. Each UAV owner is equipped with a stochastic alarm clock (Poisson-distributed), the ring of which indicates the arrival of a strategy revision opportunity. 

\textit{Mean dynamics} captures change of strategy distribution over the populations over the time. The expected change in the \textit{proportion} of UAV owners choosing strategy $m$ in population $p$ captured by the mean dynamics corresponding to the population game $F$ and revision protocol $\rho$ is defined by
\begin{equation}\label{eq-mean-dynamics}
\dot{x}^p_m=\sum_{i\in\mcal{S}^p}x^p_i\rho^p_{i,m}(\pi^p,\bm{x}^p)-x_m^p\sum_{i\in\mcal{S}^p}\rho^p_{m,i}(\pi^p,\bm{x}^p),
\end{equation}
where the first term captures switches to strategy $m$ from other strategies, whereas the second term captures switches to other strategies from strategy $i$. 

Next, we specify $\rho$ by two general categories, imitative protocol, exemplified by replicator dynamics, and direct protocol, represented by the Smith dynamics. The difference between replicator dynamics and Smith dynamics is the amount of payoff information required for \textit{one} UAV strategy adaption: replicator dynamics requires only one piece of information from the random opponent, whereas the Smith dynamics requires the $S^p$ payoff information.

\subsubsection{Imitative Protocols and Dynamics \cite{niyatoDynamicsNetworkSelection2009}}
This scheme describes a UAV owner playing strategy $m$ who receives a revision opportunity chooses an opponent randomly and observes the opponent's strategy. If the opponent is playing strategy $i$, then the owner switches from strategy $m$ to $i$ with probability proportional to some factor $r^p_{mi}$. Note that the factor $x^p_m$ needs not be observed by the owner. In this case, the imitative revision protocol is defined as
$\rho^p_{m,i}(\pi^p, x)=x^p_{i} r^p_{mi}(\pi^p, \bm{x})$, 
where $x^p_{i}$ represents the chance of meeting an opponent playing strategy $i$. 

\textit{Pairwise Proportional Imitation} is an example of an imitation protocol, in which an UAV imitates the opponent's behavior only if the opponent's payoff is higher than his own, with probability proportional to the payoff difference: $
\rho^p_{m,i}(\pi^p, x)=x^p_{i}\left[\pi^p_{i}-\pi^p_{m}\right]_{+}$. 
The mean dynamics \eqref{eq-mean-dynamics} with this revision protocol defines the best known dynamic in evolutionary game theory, \textit{replicator dynamics}, given as follows:
\begin{align}
\dot{x}^p_{m} &=\sum_{i \in \mcal{S}^p} x^p_{i} \rho^p_{im}(\pi^p, \bm{x}^p)-x^p_{m} \sum_{i \in  \mcal{S}^p} \rho^p_{m i}(\pi^p, \bm{x}^p) \nonumber \\
&=\sum_{i \in  \mcal{S}^p} x^p_{i} x^p_{m}\left[\pi^p-\pi^p_i\right]_{+}-x^p_{m} \sum_{i \in  \mcal{S}^p} x_{i}^p\left[\pi^p_i-\pi^p\right]_{+} \nonumber \\
&=x^p_{m} \sum_{i \in  \mcal{S}^p} x^p_{i}\left(\pi^p-\pi^p_i\right) =x^p_{m}\left(\pi^p_{m}-\bar{\pi}^p\right),
\label{eq-replicator-dynamics}
\end{align}
where ${\bar{\pi}}^p=\sum_{m\in S^p}x^p_m\pi^p_m$ is the \textit{average payoff} of population $p$, given the social state $\bm{x}$,

\subsubsection{Direct Protocol and Dynamics \cite{sandholmPopulationGamesEvolutionary2011}}
This scheme describes the situation that an UAV device owner in population $p$ can receive a full picture of the payoff information ($S^p$ payoff) from a centralized controller, e.g., by requesting the payoff information from the BS (which connects to different UAVs) or checking the bulletin board, a place to publish virtual task \cite{duanMetaverseSocialGood2021}, in the Metaverse platform. Therefore, the UAV owner's conditional switch rate is directly dependent on the excess payoffs between two strategies. This is called 
a \textit{pairwise comparison protocol} defined by $\rho^p_{m,i}(\pi^p)=[\pi_i-\pi_m]_+$,
representing the case when a UAV owner, who has a chance to switch strategy, randomly chooses an alternative strategy with higher payoff than the current one, with a probability proportional to the difference between the two payoffs. The mean dynamic \eqref{eq-mean-dynamics} with this revision protocol is called the \textit{Smith dynamics} defined as follows:
\begin{align}\label{eq-smith}
	\dot{x}^p_m &=  \sum_{j\in\mcal{S}^p}x^p_j \rho^p_{jm}- x^p_m \sum_{j\in\mcal{S}^p}\rho^p_{mj} \nonumber  \\
	&= \sum_{j\in\mcal{S}^p}x^p_j[\pi^p_m -\pi^p_j ]_+ - x^p_m  \sum_{j\in\mcal{S}^p}[\pi^p_j -\pi^p_m ]_+.
\end{align}

\subsubsection{Hybrid Protocol for Heterogeneous Multi-population \cite{sandholmPopulationGamesEvolutionary2011}}
Let $\mcal{K}=\{1, \ldots, k,\ldots, K\}$ denote a set of revision protocols available to the society, and $\rho^{p,k}(\pi^p,\bm{x}^p)$ refers to the conditional switch rate that an UAV owner in population $p$ has under revision protocol $k$. Let $\alpha^{p,k}(k\in\mcal{K})$ satisfying $\sum_{k\in\mcal{K}} \alpha^{p,k}  = 1$ denote the probability of a $p$-population member using revision protocol $\rho^{p,k}$. 
Then, the behavior of an owner in population $p$ can be described by a \textit{hybrid revision protocol} $\rho^{p,H}$, which is expressed as $\rho^{p}= \sum_{k\in\mcal{K}}\alpha^{p,k}\rho^{p,k}$. Note that mean dynamics \eqref{eq-mean-dynamics} are linear in conditional switch rates, 
The mean dynamics with the hybrid revision protocol can be expressed as: 
\begin{equation*}\label{key}
	\dot{x}^p_m=  \sum_{i\in\mcal{S}^p}\sum_{k\in\mcal{K}}x^p_i \alpha^{p,k} \rho^{p,k}_{i,m}-
	x_m^p  \sum_{i\in\mcal{S}^p}\sum_{k\in\mcal{K}} \alpha^{p,k} \rho^{p,k}_{m,i}.
\end{equation*}

The hybrid protocol is versatile and suitable for different situations, e.g., including other protocols as introduced in \cite{sandholmPopulationGamesEvolutionary2011}. Unless otherwise stated, in the following, the hybrid protocol consists of only replicator dynamics and Smith dynamics. We use $\alpha^{p,1}$ to refer to the population $p$'s probability of switching to the centralized communication approach (observing $S^p$ payoffs) and $1-\alpha^{p,1}$ for pairwise approach (observing $1$ opponent's payoff). In other words, when adjusting strategy, a $p$-population UAV owner's behavior can be modeled by Smith and replicator dynamics with probability $\alpha^{p,1}$ and $1-\alpha^{p,1}$, respectively.

\section{Simulations Results}\label{sec-simulation}
Unless stated otherwise, we consider the problem formulated with $3$ populations selecting among $3$ VSPs to assist data sync. Population size $N^p$ lies in $[50,250]$ with traversal distance $l^p_m$ in $[0.3,1]$ km. The length of sensing route $D_m$ lies in $[1,1.8]$ km, whereas the reward pool $R_m$ lies in $[1000,2000]$. Following \cite{limFederatedLearningUAVEnabled2021,zhangPredictiveDeploymentUAV2021}, values of speed parameters $v^p,u^p$ and power parameters $\eta^p_1,\eta^p_2$ are in the range $[3,5]$ m/s and $[16,20]$ W, respectively. 
Sensing data quality $b^p_m$ is in the range $[1,5]$, given the accuracy of sensors can be scored and calibrated in this range, whereas the unit energy cost $\zeta$ is $0.001$ 
\$/Joule. The default probabilities of adopting Smith protocol is $\alpha^1=[\alpha^{p,1}]_{p=1,2,3}=[0.2,0.3,0]$. 

\begin{figure*}[]
\begin{minipage}[b]{0.3\linewidth}
\centering
\includegraphics[width=\linewidth]{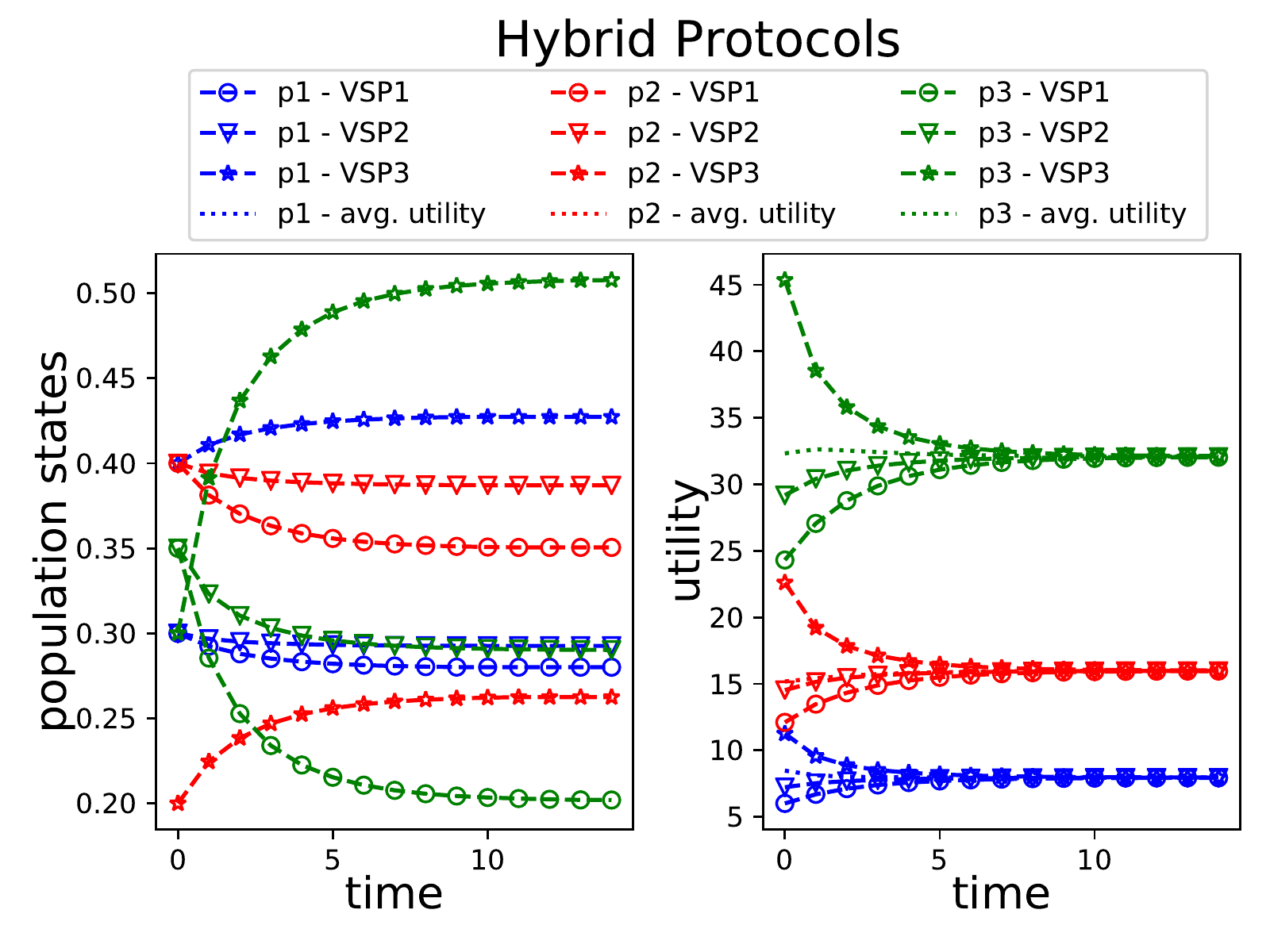}
\caption{Population states and utilities vs. time. Plot every $10$ steps. }
\label{fig:hybrid-converge}
\end{minipage} \quad
\begin{minipage}[b]{0.45\linewidth}
\centering
\includegraphics[width=\linewidth]{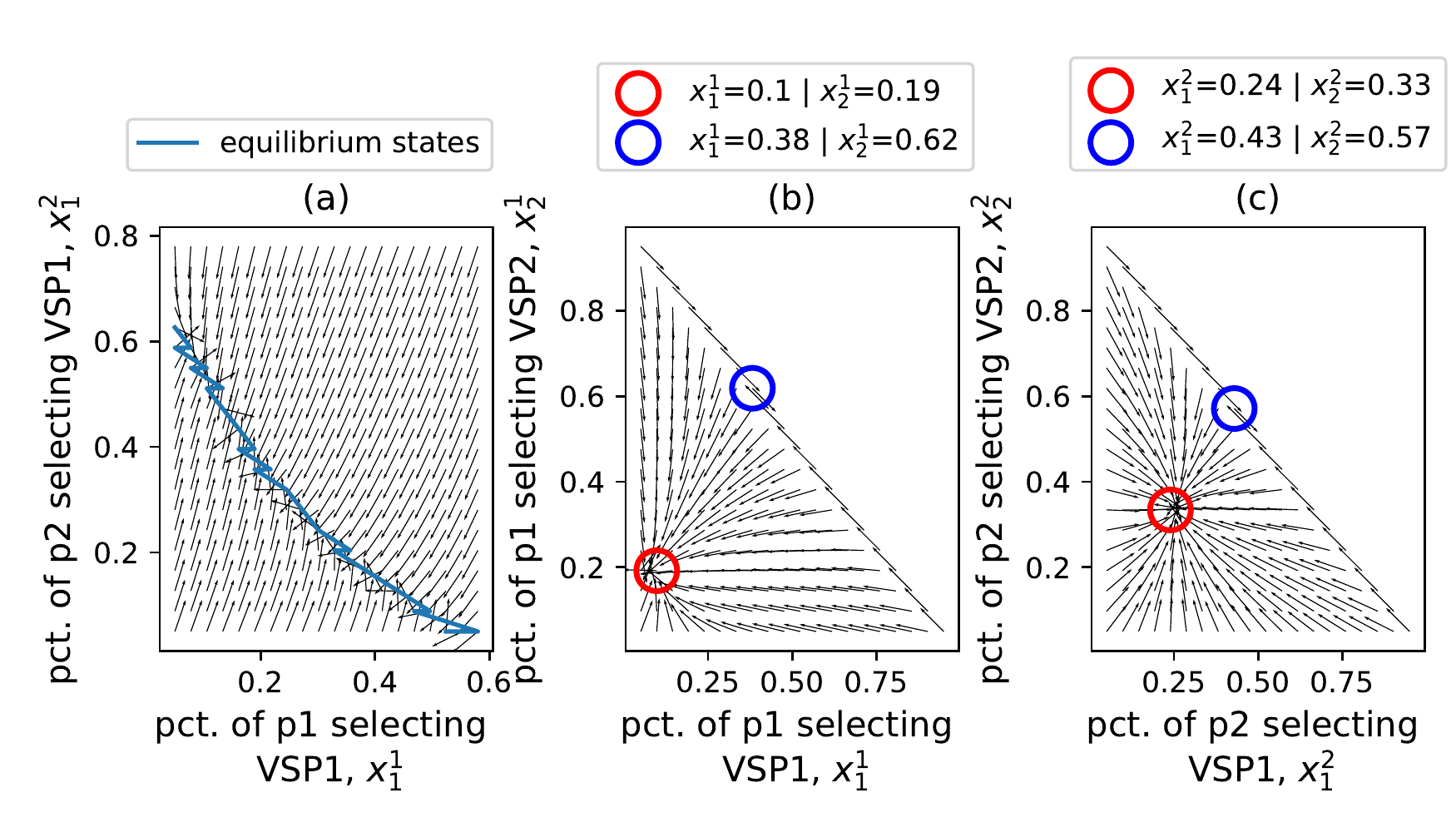}
\caption{Direction field. (pct. is short for percentage)}
\label{fig:vf}
\end{minipage} \quad
\begin{minipage}[b]{0.2\linewidth}
\centering
\includegraphics[width=0.68\linewidth]{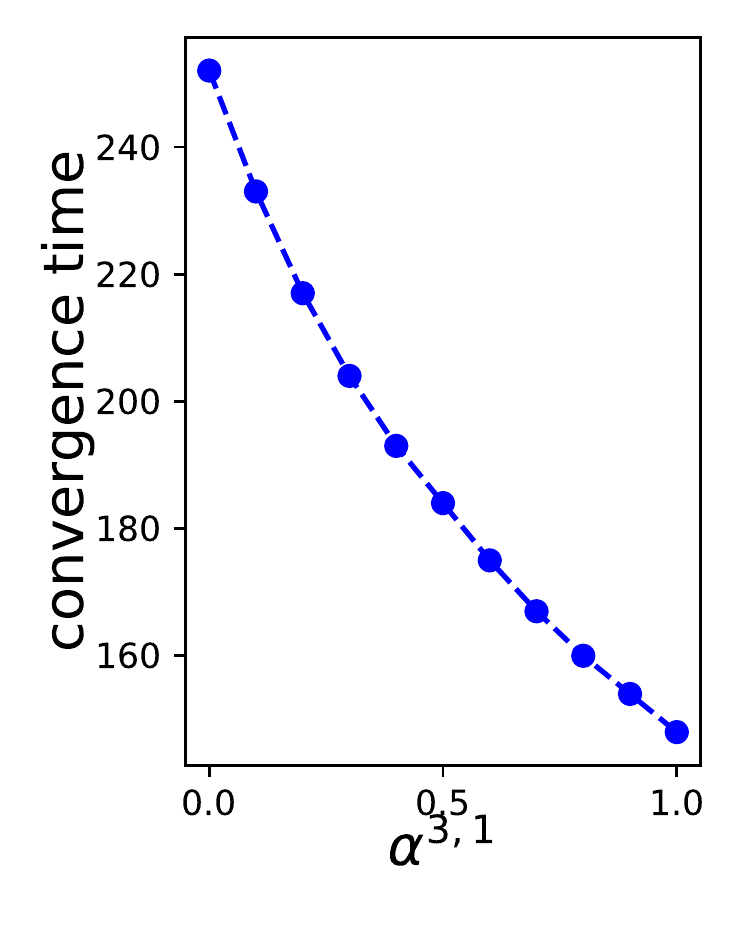}
\caption{Convergence time vs. $\alpha^{3,1}$}
\label{fig:hybridlevel}
\end{minipage}
\end{figure*}

\paragraph{Existence of the Equilibrium Strategies}
\cref{fig:hybrid-converge} considers the hypothetical 
scenarios in which owners in different populations have various probabilities of switching to the centralized communication protocol, e.g., by checking the payoffs on the bulletin board in the Metaverse platform. The initial society states (strategy distribution) is given as $\bm{x}_0=[[0.3,0.3,0.4],[0.4,0.4,0.2],[0.35,0.35,0.3]]$. The left subfigure shows that strategy distribution can be stationary (i.e., convergent at equilibrium states) at which the payoffs received of selecting any VSP are the same (as shown in the right subfigure). Thus, the owners in each population have no incentive to make any more adjustment to their strategies away from the equilibrium point. A population-$3$ UAV receives higher average payoff than that of population $2$ and $3$ due to higher data quality.

\paragraph{Stability of the Equilibrium Strategies}
\cref{fig:vf} shows the stability of the equilibrium strategies using direction field, a way of graphically representing the evolution direction (arrows in the figure) for the population states. For visualization purposes, population $3$ is removed from the default setting, and the probability of selecting region $3$ is fixed. The evolution direction, change in $(x^1_1,x^1_2,x^2_1,x^2_2)$ at a time instant, is four-dimensional. \cref{fig:vf}(a) shows the case when $x^1_1$ and $x^2_1$ are varied, and the evolution direction along $x^1_2$ and $x^2_2$ are hidden. Initial population states will follow the arrows and evolve towards the equilibrium states, shown as a blue line in the figure. 
\cref{fig:vf}(b) shows that when $x^2_1,x^2_2$ are fixed and $x^1_1$ and $x^1_2$ are varied, a mixed strategy equilibrium state $(0.1,0.19)$ can be obtained (red circle). Note that the blue circle $(0.38,0.62)$ is also an equilibrium state, at which the chance of selecting region $3$ is zero (i.e., extinction of strategy $3$ in population $1$). Finally, \cref{fig:vf}(c) shows the case when states of population $1$, ($x^1_1,x^1_2$), are fixed and those of population $2$, ($x^2_1,x^2_2$), are varied, the equilibrium point $(0.24,0.33)$ can be achieved. $(0.43,57)$ is also an equilibrium point at which strategy $3$ is extinct. 

\paragraph{Impact of Smith Protocol on the Convergence Time}
\cref{fig:hybridlevel} shows that the impact of adopting Smith protocol on the convergence time by fixing the probability of adopting the Smith protocol in population $1$ and $2$ and varying that in population $3$, $\alpha^{3,1}$, within $[0,1]$ in increments of $0.1$. The convergence time can be defined as the number of iterations after which the utilities among strategies in a population are similar, given the UAV owners in that population will have no incentive to make any modification, and the equilibrium state is reached. Mathematically, convergence time is $\min_{t}\left| \max_{i\in \mcal{S}^p}\pi^p_i(t) - \min_{i\in \mcal{S}^p}\pi^p_i(t)\right|\leq \tau$, where $\tau$ is a threshold, chosen as $0.05$ in the experiment. The figure shows that the convergence time decreases as $\alpha^{3,1}$ increases. This is because complete information of all payoffs improves the strategy adaption compared to pairwise imitation in which only the opponent's payoff is known. 

\section{Conclusion}
In this paper, we studied an IoT-assisted Metaverse sync problem in which the populations of UAV devices can autonomously select VSPs to work for. Given a large number of IoT devices, hardly-achieved full rationale, and potential different in-population communication protocols, we proposed hybrid evolutionary dynamics for a heterogeneous multi-population game. The equilibrium strategy is experimentally demonstrated to be existed and stable. In future works, we plan to incorporate spatial correlation of the sensing data into the utility models.

\bibliographystyle{IEEEtran}
\bibliography{all.bib}

\end{document}